\documentclass{article}

\usepackage{graphicx}
\usepackage[english]{babel}

\usepackage{amssymb}
\usepackage{hhline}
\usepackage{epsfig}
\usepackage{amsmath}
\usepackage{dcolumn}
\usepackage{multirow}
\usepackage{array}
\usepackage{color}

\newcommand{\be}{\begin{equation}}
\newcommand{\ee}[1]{\label{#1} \end{equation}}
\newcommand{\ba}{\begin{eqnarray}}
\newcommand{\ea}[1]{\label{#1} \end{eqnarray}}
\newcommand{\nl}{\nonumber \\}

\title{\includegraphics[width=0.35\textwidth, height=0.35\textwidth]{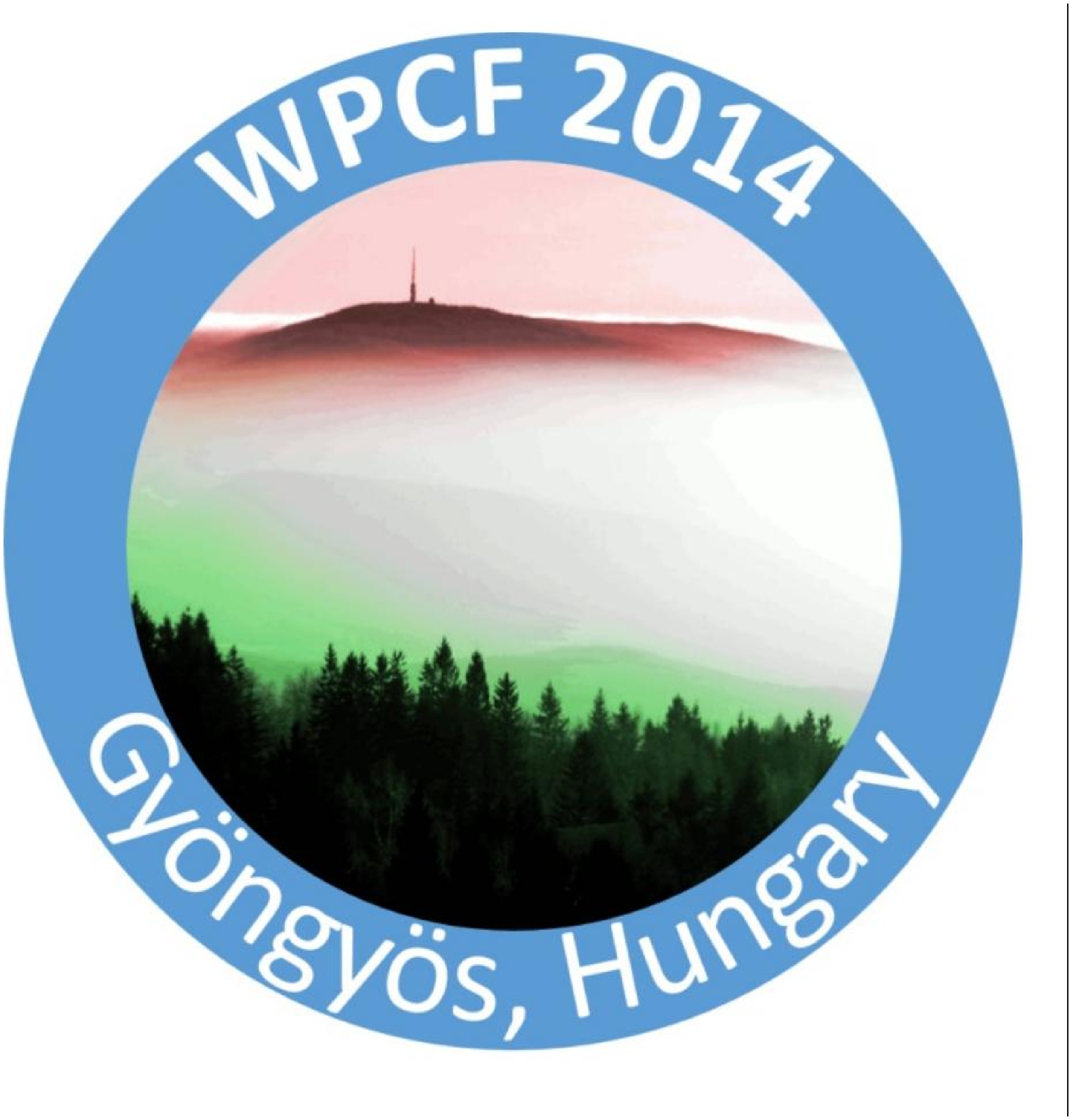}\\[1cm]
$v_2$ of charged hadrons in a`soft + hard' model\\ for PbPb collisions at $\sqrt s$ = 2.76 ATeV }
\author{{K. Urmossy$^{\,1}$, T.~S.~Bir\'o$^{\,1}$, G. G. Barnaf\"oldi$^{\,1}$ and Z. Xu$^{\,2}$,}\\[1ex]
$^1$ Wigner Research Center for Physics of the HAS,\\ 29--33 Konkoly--Thege Mikl\'os Str.\\ H-1121 Budapest, Hungary\\
$^2$ Physics Department, Brookhaven National Laboratory,\\Upton, NY 11973, USA\\}

\begin{document}

\fontfamily{lmss}\selectfont
\maketitle

\begin{abstract}
We describe transverse spectra \textit{as well as} azimuthal anisotropy ($v_2$) of charged hadrons stemming from various centrality PbPb collisions at $\sqrt s$ = 2.76 ATeV \textit{analytically} in a `soft + hard' model. In this model, we propose that hadron yields produced in heavy-ion collisions are simply the sum of yields stemming from jets (hard yields) and yields stemming from the Quark-Gluon Plasma (soft yields). The hadron spectra in both types of yields are approximated by the Tsallis distribution. It is found that the anisotropy decreases for more central collisions.
\end{abstract}

\section{Introduction}
\label{sec:intro}

Because of its short lifetime, the only way to examine the Quark-Gluon Plasma (QGP) formed in ultra-relativistic heavy-ion collisions (HIC), is looking at the particles stemming from it. Spectra, angular correlations and their dependence on the circumstances of the collision can then be studied. These distributions are effected by hadron yields stemming not only from the QGP (we refer to as `soft' yields), but also from jets (we call `hard' yields).

As a first approximation, we make out hadron spectra in HICs as
\be
p^0\frac{dN}{d^3\mathbf{p}} =  p^0\frac{dN}{d^3\mathbf{p}}^{\textrm{hard}} + p^0\frac{dN}{d^3\mathbf{p}}^{\textrm{soft}} \;,
\ee{eq1}
where we describe both types of yields by a Tsallis distribution with different parameters for the following reasons.
\begin{itemize}
\item[]\textbf{Hard yields:} On the experimental side, the Tsallis distribution describes measured transverse spectra of charged and identified hadrons in proton-proton collisions \cite{bib:phenixPP}--\cite{bib:Wong2}. On the theoretical side, the Tsallis distribution provides a reasonably good approximation for the transverse spectra of charged pions stemming from pp collisions \cite{bib:Wong1,bib:Wong2}, and central as well as peripheral PbPb collisions obtained via perturbative quantum chromodynamics (pQCD) improved parton model calculations for transverse momenta $p_T\gtrsim$ 4--6 GeV/c \cite{bib:UKshAA}.
\item[]\textbf{Soft yields:} The Tsallis distribution has been widely used for the description of hadron yields stemming from the QGP \cite{bib:Wilk2}--\cite{bib:Wilk7}. However, in those models, the hard part of the spectrum has not been subtracted. For the emergence of the Tsallis distribution in the soft part of the spectrum, there is a chance to bring statistical arguments based on non-extensive thermodynamics \cite{bib:BiroEPJA40,bib:BiroJako}, or on super-statistics \cite{bib:UKppNdep,bib:Wilk3,bib:Wilk4,bib:Wilk5,bib:Wilk6,bib:Wilk7,bib:UKppFF,bib:UKeeFF}.\\
We note that transverse spectra and $v_2$ of various identified hadrons measured at RHIC energy have been described by a similar model \cite{bib:Tang1,bib:Tang2,bib:Shao}. In that model, spectra measured in $pp$ collisions have been used as hard yields, and it has been conjectured that hard yields are suppressed at low $p_T$.
\end{itemize}

In Sec.~\ref{sec:vn}, analytic formulas are decuced for the hadron spectrum and $v_2$. Sec.~\ref{sec:fitspec} contains fits to charged hadron spectra and $v_2$ measured in various centrality PbPb collisions at $\sqrt s$ = 2.76 ATeV by the CMS \cite{bib:cmsdNdpT,bib:cmsv2} and the ALICE \cite{bib:ALICEdNdpT} collaborations. Summary is given in Sec.~\ref{sec:conc}.

\section{Transverse Spectrum and $v_n$}
\label{sec:vn}

In statistical models, we obtain the transverse spectrum as a sum of hadrons with momentum $p^{\mu}$, coming from sources flying with velocities $u^{\mu}$ as
\be
\left.p^0\,\frac{dN}{d^3p}\right|_{y=0} = \int \limits_{-\infty}^{+\infty} d\zeta \int \limits_0^{2\pi} d\alpha \, f\big[ u_\mu p^\mu \big] \;.
\ee{eq2}
Here, $\alpha$ is the azimuth angle and $\zeta = \frac{1}{2}\ln[(t+z)/(t-z)]$. We parametrize hadron momenta as
\be
p^{\,\mu} = (m_T\cosh y,m_T\sinh y,p_T \cos\varphi, p_T\sin\varphi)\;,
\ee{eq3}
with $y = \frac{1}{2}\ln[(p^0+p^z)/(p^0-p^z)]$ and $\varphi$ being the azimuth angle of the hadron momentum. We parametrize the flow as
\be
u^{\,\mu} = (\gamma\cosh\zeta,\gamma\sinh\zeta,\gamma v\cos\alpha, \gamma v \sin\alpha)\;,
\ee{eq4}
with $\gamma = 1/\sqrt{1-v^2}$, and assume that $v$ depends only on $\alpha$. Though, it is assumed that in each source, the momentum  distribution of hadrons $f$ is a function of the co-moving energy
\be
\left. u_\mu p^\mu \right|_{y=0} = \gamma \big[m_T\cosh\zeta - v p_T \cos(\varphi-\alpha) \big]\;,
\ee{eq5}
the sources may be fireballs \cite{bib:OldBoys1}--\cite{bib:OldBoys3}, clusters \cite{bib:Wibig2,bib:Wibig,bib:Wibig3,bib:Becattini11,bib:Liu1} or even jets \cite{bib:UKppFF,bib:UKeeFF}.

We write the transverse flow as a series,
\be
v(\alpha)= v_0 + \sum\limits_{m=1}^\infty \delta v_m \cos(m\alpha) \equiv v_0 + \delta v(\alpha) \;,
\ee{eq6}
and suppose that $\delta v(\alpha) << 1$. We use the Taylor expansion 
\be
\left.f\big[u_\mu p^\mu \big]\right|_{y=0} = \sum_{m=0}^\infty \frac{ \left[\delta v(\alpha)\right]^m}{m!} \frac{\partial^m}{\partial v_0^m}\, \left.f\big[ u_\mu p^\mu \big]\right|_{\,y\,=\,0}^{\,v(\alpha)\,=\,v_0}\;,
\ee{eq7}
and keep only the leading non-vanishing terms in $\delta v(\alpha)$.

Provided that $f$ is a rapidly decreasing function, we approximate integrals with respect to $\zeta$ and $\varphi$ by the maximal value of the integrands times the integration interval. Thus, the $\varphi$ integrated transverse spectrum becomes
\ba
\left. \frac{dN}{2\pi p_T dp_T dy} \right|_{\,y\,=\,0} &=& \int \limits_0^{2\pi} \frac{d\varphi}{2\pi}  \left.p^0\,\dfrac{dN}{d^3p}\right|_{\,y\,=\,0} \;=\; 
\sum_{m=0}^\infty \frac{a_m}{m!} \frac{\partial^m}{\partial v_0^m}\, f\big[E(v_0)\big] \;\approx\; \nl
&\approx& f\big[E(v_0)\big] \;+\; \mathcal{O}\left(\delta v^2\right)\;,
\ea{eq8}
with $E(v_0) = \gamma_0 (m_T - v_0 p_T)$ and $a_m = \int\limits_0^{2\pi} d\alpha \,\left[\delta v(\alpha)\right]^m$. Similarly, the azimuthal anisotropy becomes
\ba
v_n = \frac{\int \limits_0^{2\pi} d\varphi \cos(n\varphi) \, \left.p^0\,\dfrac{dN}{d^3p}\right|_{y=0}}{ \int \limits_0^{2\pi} d\varphi \, \left.p^0\,\dfrac{dN}{d^3p}\right|_{y=0} } 
\approx \frac{\delta v_n \gamma^3_0}{2} \frac{(v_0\, m_T - p_T) f'\big[E(v_0)\big]}{f\big[E(v_0)\big]} + \mathcal{O}\left(\delta v^2\right)
\ea{eq9}
with $\delta v_n$ defined in Eq.~(\ref{eq6}).

For example, in the case of the Boltzmann-distribution $f \sim \exp\big[-\beta E(v_0)\big]$, the anisotropy is
\be
v^{\textrm{BG}}_n \approx \frac{\delta v_n \,\beta\, \gamma^3_0}{2} (p_T - v_0\, m_T) \; +\; \mathcal{O}\left(\delta v^2\right)\;.
\ee{eq10}
Thus, $v^{\textrm{BG}}_n \propto p_T$ if $p_T>>m$.

In the case of the Tsallis distribution, $f \sim [1 + (q-1)\,\beta\, E(v_0)]^{-1/(q-1)}$, the anisotropy
\be
v^{\textrm{TS}}_n \approx \frac{\delta v_n \,\beta\, \gamma^3_0}{2} \frac{p_T - v_0\, m_T}{1 + (q-1)\,\beta\,\gamma_0(m_T - v_0\,p_T)} \; +\; \mathcal{O}\left(\delta v^2\right) \;.
\ee{eq11}
Thus, $v^{\textrm{TS}}_n$ saturates when $(q-1)\,\beta\,\gamma_0 (1-v_0) p_T >> 1$.

\section{Fits to the spectrum and $v_2$ of charged had\-rons in PbPb collisions at $\sqrt{s_{NN}}$ = 2.76 ATeV}
\label{sec:fitspec}
\begin{figure}
\begin{center}
\includegraphics[width=0.45\textwidth, height=0.283\textheight]{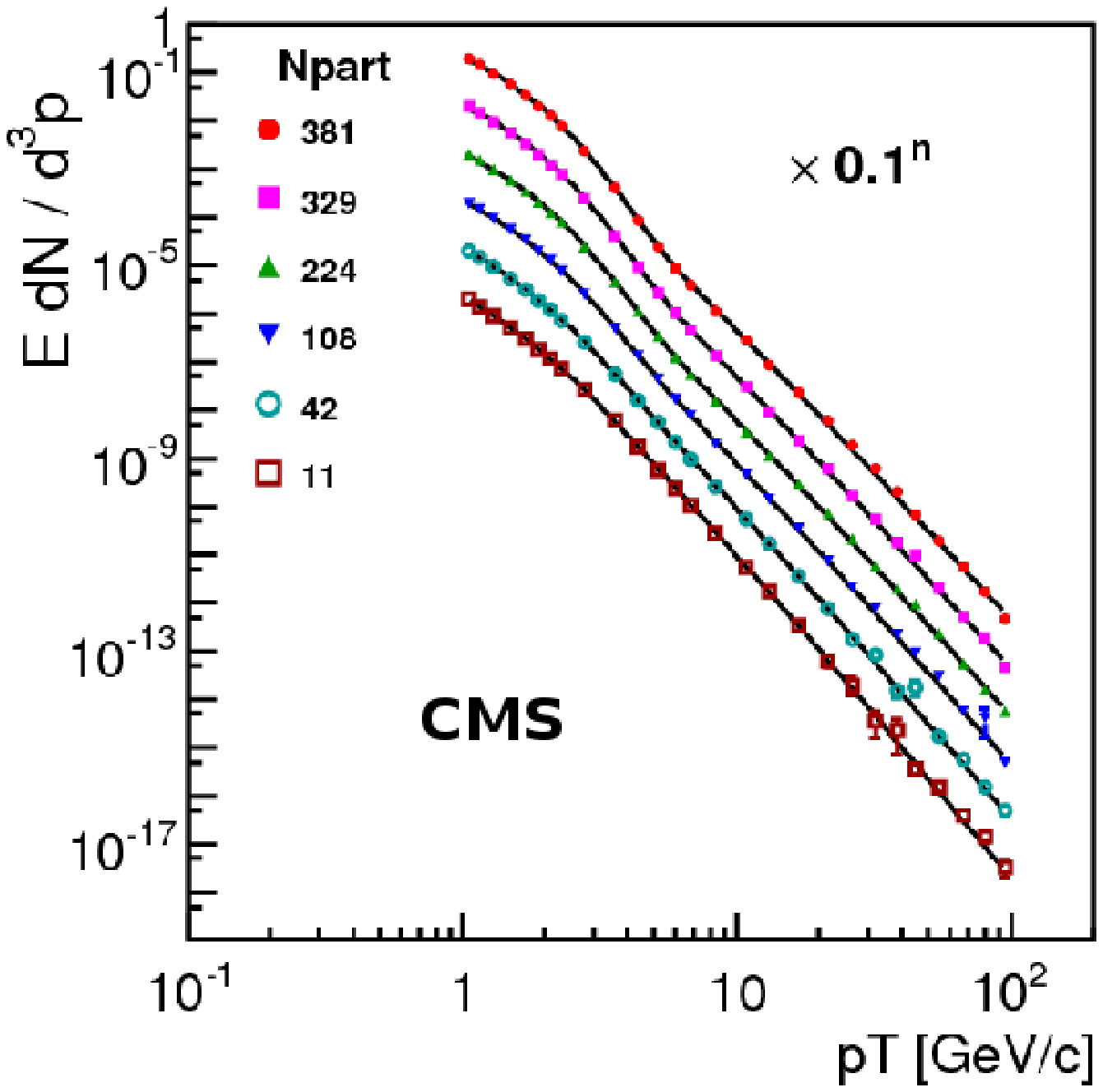}
\includegraphics[width=0.45\textwidth]{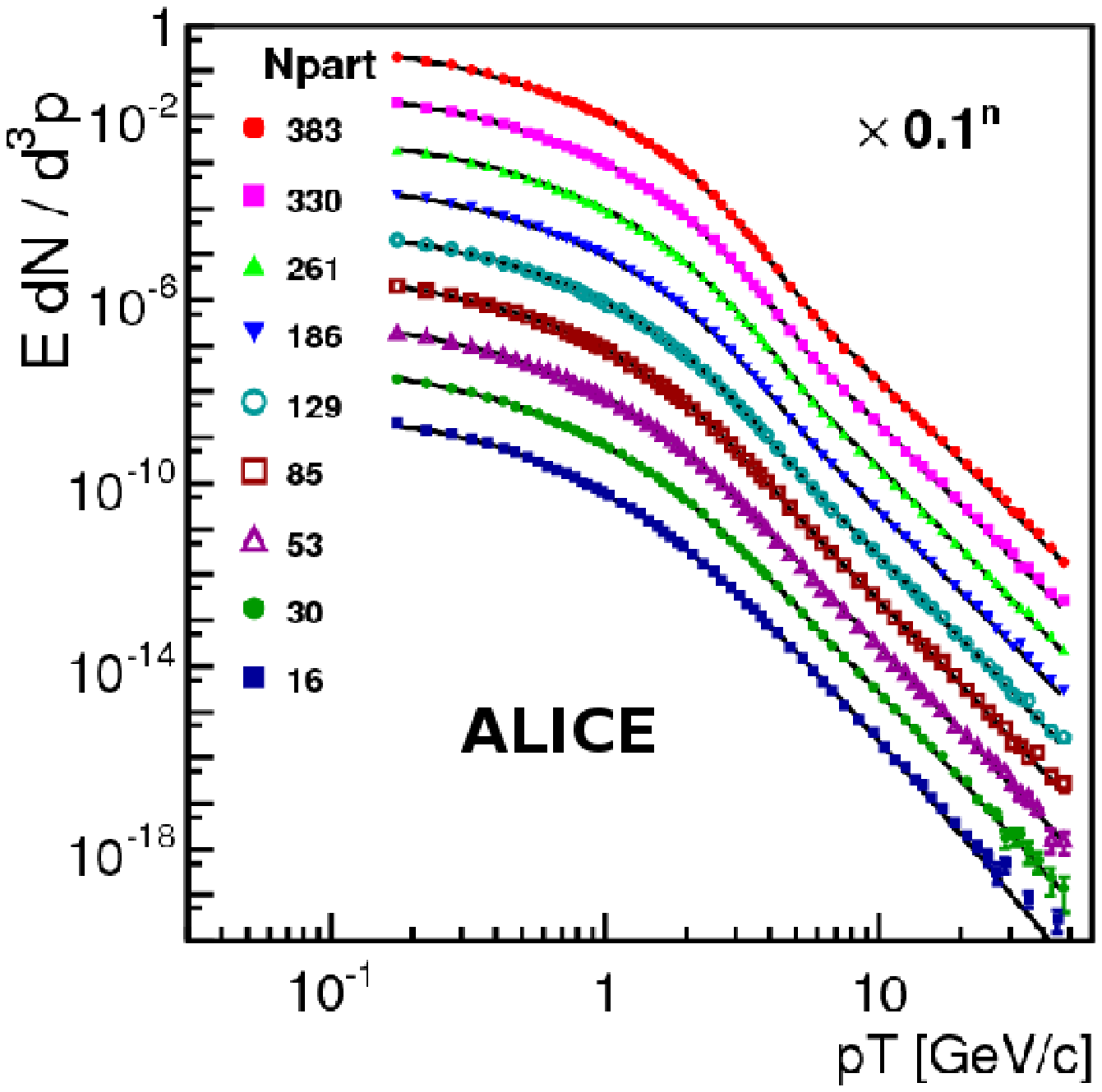}
\includegraphics[width=0.45\textwidth, height=0.3\textheight]{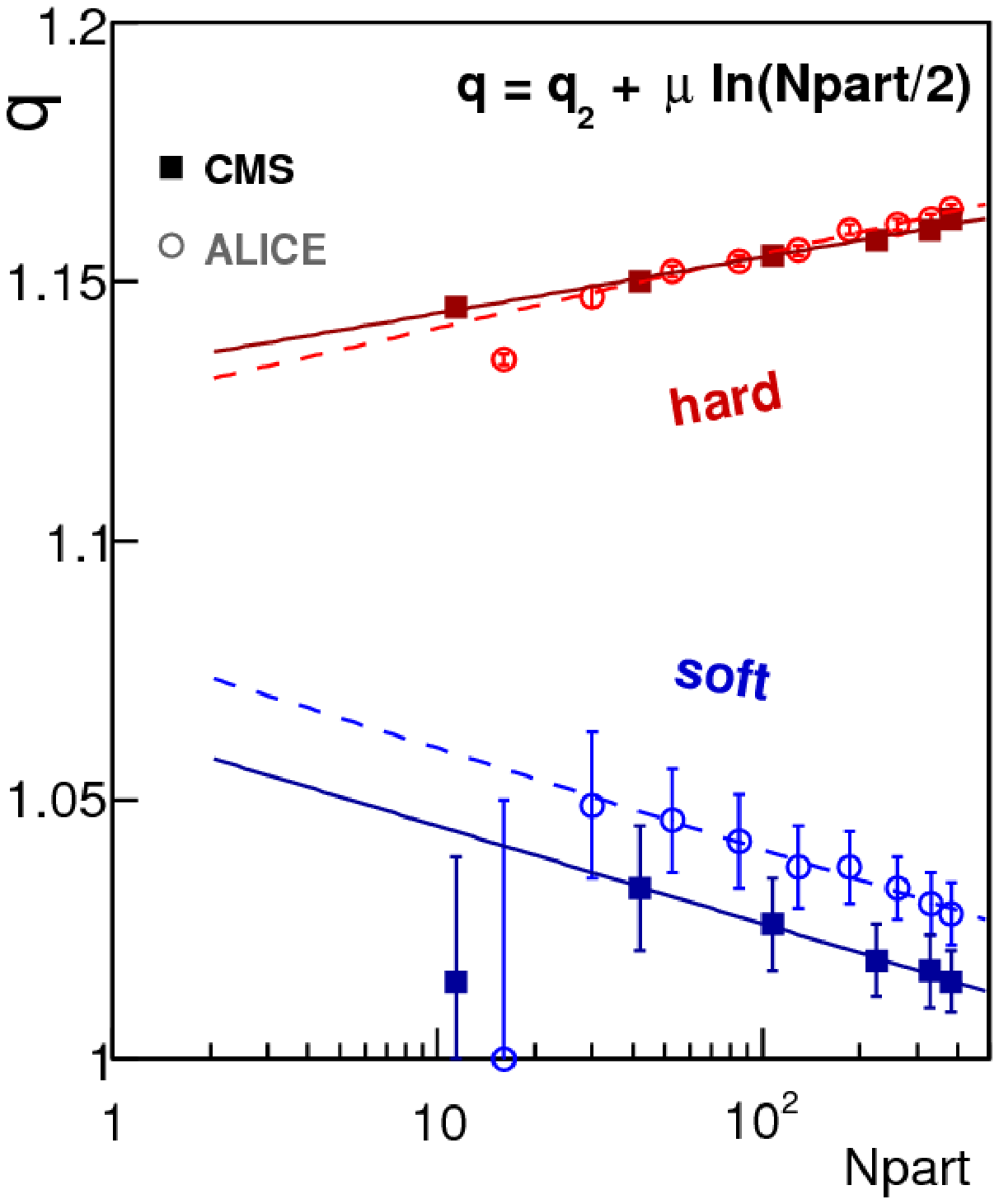}
\includegraphics[width=0.45\textwidth, height=0.303\textheight]{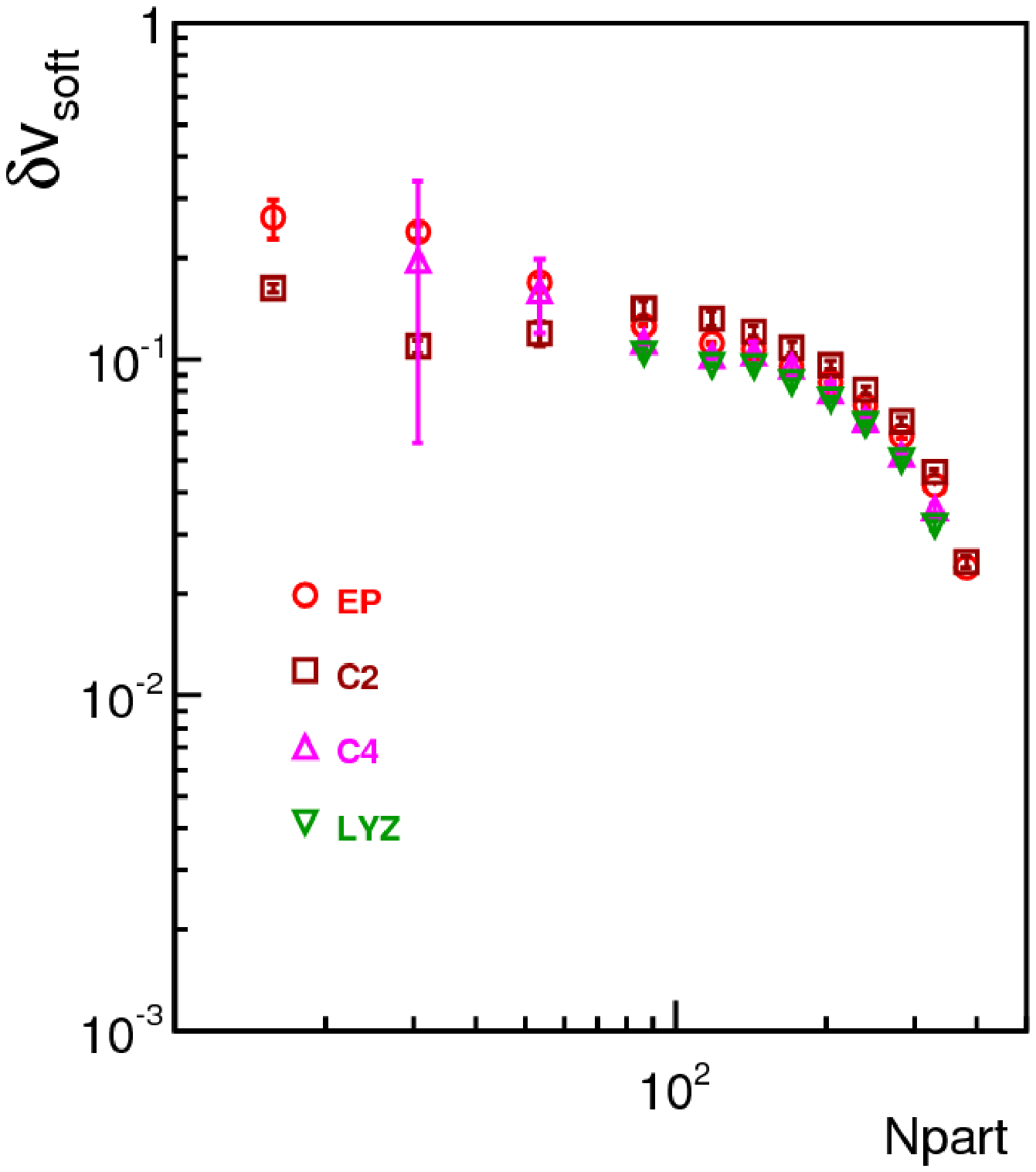}
\end{center}
\caption{\textbf{Top,} transverse spectra of charged hadrons stemming from various centrality PbPb collisions at $\sqrt s$ = 2.76 ATeV measured by the CMS \cite{bib:cmsdNdpT} (\textbf{left}) and ALICE \cite{bib:ALICEdNdpT} (\textbf{right}) Collaborations. Curves are fits of Eq.~(\ref{eq12}). \textbf{Bottom,} centrality dependence of the fitted $q$ (\textbf{left}) and $\delta v_{soft}$ (\textbf{right}) parameters. Straight lines are in Eq.~(\ref{eq15}) with parameters enlisted in \cite{bib:UKshAA}. \label{fig:dNdpT}}
\end{figure}
As conjectured in Sec.~\ref{sec:intro}, we make out the transverse spectrum of charged hadrons by the sum of hard and soft yields
\be
\left. \frac{dN}{2\pi p_T dp_T dy} \right|_{y=0} = \sum_i A_i \left[1 + \frac{(q_i-1)}{T_i}[\gamma_i(m_T-v_i p_T) - m] \right]^{-1/(q_i-1)}\;,
\ee{eq12}
($i$ = soft or hard) where both contributions are assumed to be Tsallis-distributions. These yields have maxima at $p^{max}_{T,\,i} = \gamma_i\, m\, v_i$. As long as these maxima are below the measurement range, which is the case in this analysis, the isotropic part of the transverse flow, $v_i$ (denoted by $v_0$ in Eq.~(\ref{eq6}) in Sec.~\ref{sec:vn}) cannot be determined accurately. As the dominant part of charged hadrons consists of pions, the argument in Eqs.~(\ref{eq12}) may be approximated by $[\gamma_i(m_T - v_i p_T) -m]/T_i \approx p_T/T^{Dopp}_i$ with the Doppler-shifted parameters
\be
T^{Dopp}_i = T_i\,\sqrt{\frac{1+v_i}{1-v_i}} \;.
\ee{eq14}

As can be seen in the top panels of Fig.~\ref{fig:dNdpT}, Eq.~(\ref{eq12}) describes CMS \cite{bib:cmsdNdpT} and ALICE \cite{bib:ALICEdNdpT} data on transverse spectra of charged hadrons stemming from PbPb collisions of various centralities. Fitted parameters are enlisted in \cite{bib:UKshAA} and shown in the bottom panels of Fig.~\ref{fig:dNdpT}. The dependence of the $q$ and $T^{Dopp}$ parameters of the soft and hard yields on the event centrality (number of participating nucleons $N_{part}$) can be fitted by
\ba
q_i &=& q_{2,\,i} + \mu_i \ln(N_{part}/2) \;,\nl
T^{Dopp}_i &=& T_{1,\,i} + \tau_i \ln(N_{part}) \;.
\ea{eq15}
Though the actual value of the transverse flow velocity cannot be determined in this model from the spectra of charged hadrons, it may be guessed using the value of the QGP-hadronic matter transition temperature obtained from lattice-QCD calculations. As the values of fitted $T^{Dopp}_{soft}$ scatter around 340 MeV, in case of a flow velocity of $v_{soft} \approx$  0.6, the real $T_{soft}$ values would scatter around 170 MeV, which is close to the lattice result obtained e.g. in \cite{bib:lQCD}.

While the tendencies of how fit parameters depend on $N_{part}$ are similar, they are not the same within errors in the case of CMS \cite{bib:cmsdNdpT} and ALICE \cite{bib:ALICEdNdpT} measurements. It is to be noted that in \cite{bib:ALICEdNdpT}, centrality is determined using the distribution of hits in the VZERO detector, which has a rapidity coverage of 2.8 $\leq\eta\leq$ 5.1 and -3.7 $\leq\eta\leq$ -1.7. In the meanwhile, in \cite{bib:cmsdNdpT,bib:cmsv2}, the collision event centrality is determined from the event-by-event total energy deposition in both Hadron Forward calorimeters having rapidity coverage of 2.9 $\leq|\eta|\leq$ 5.2.
\begin{figure}
\begin{center}
\includegraphics[width=0.45\textwidth]{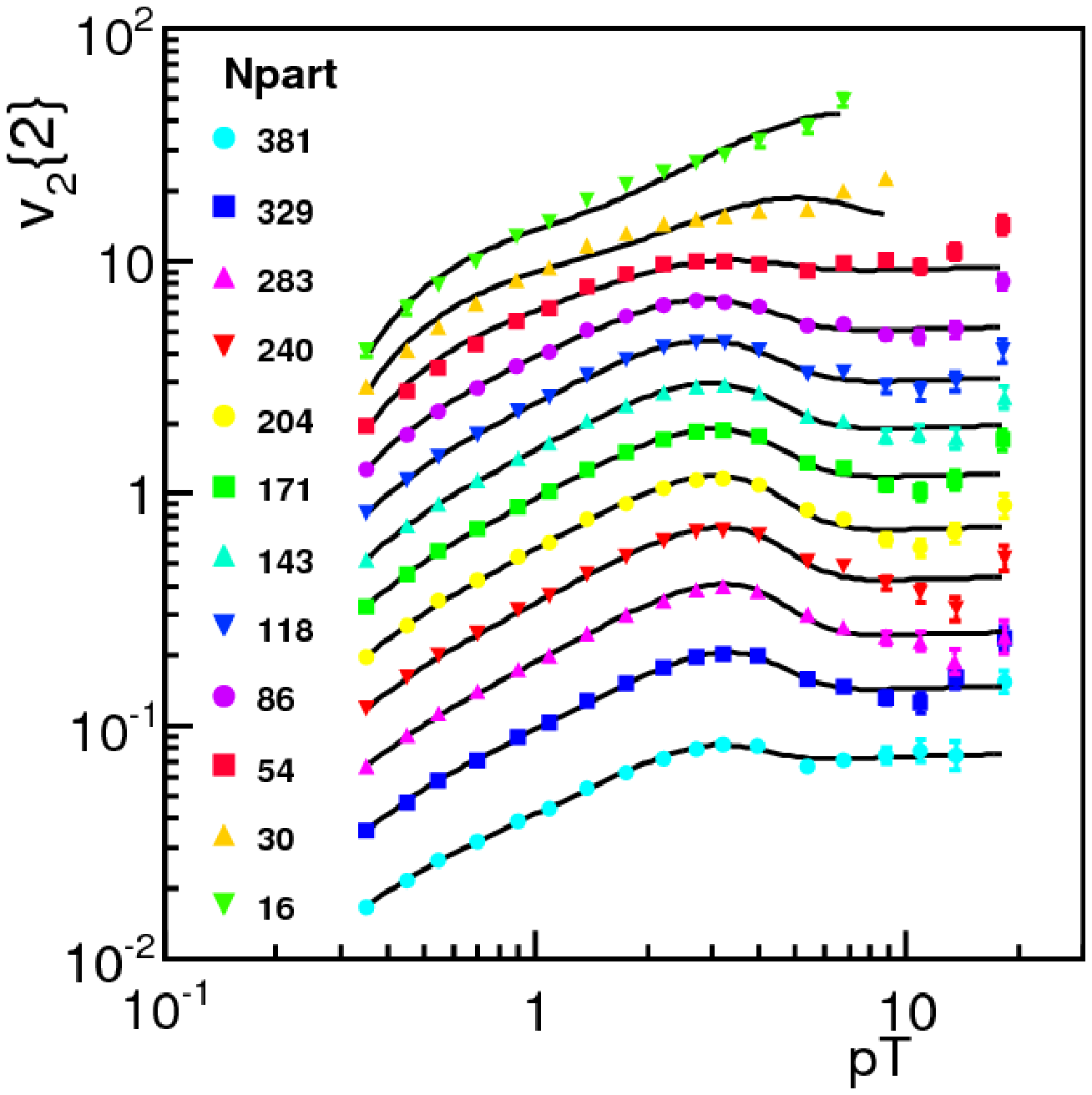} 
\includegraphics[width=0.45\textwidth]{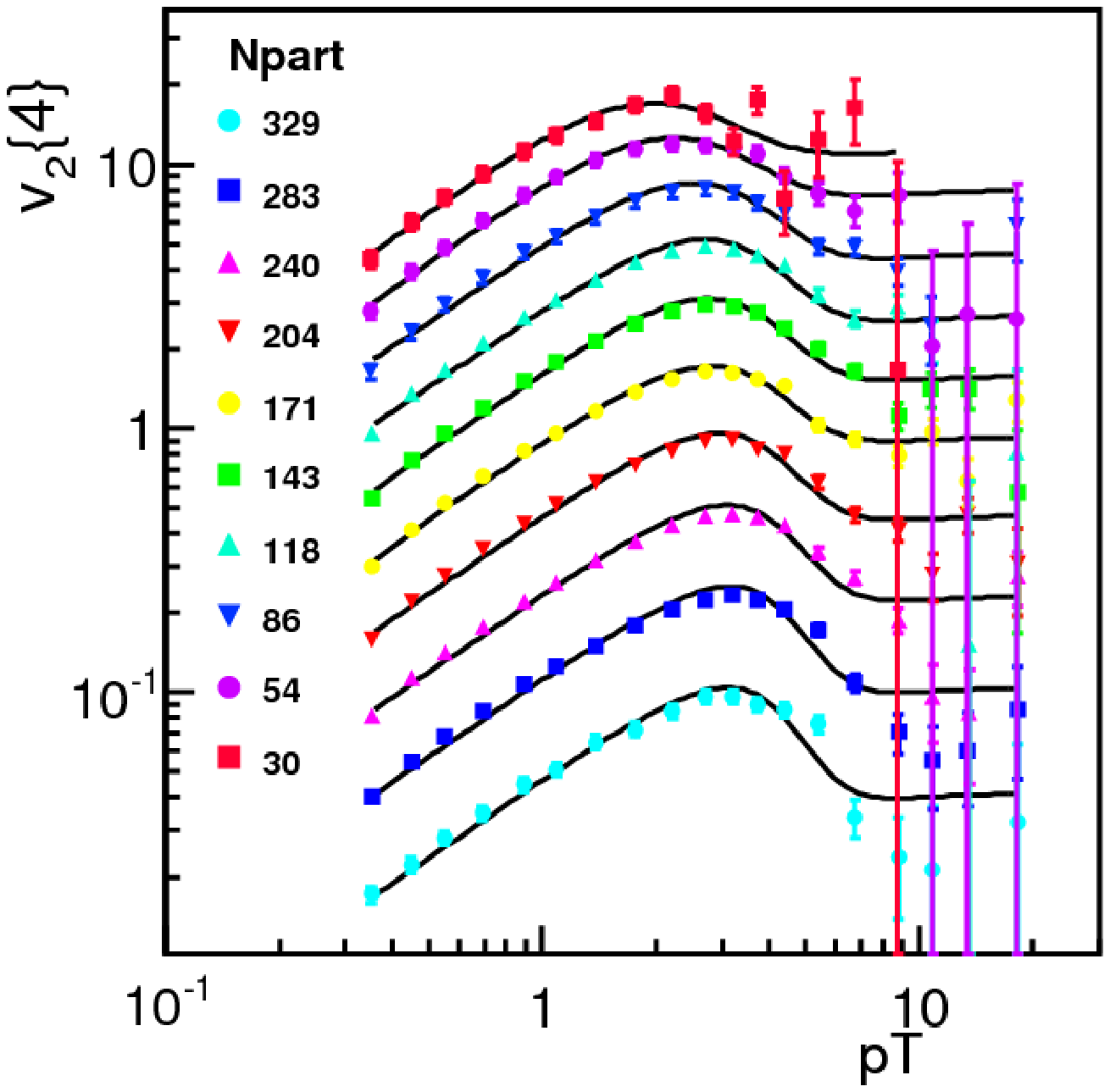} 
\includegraphics[width=0.45\textwidth]{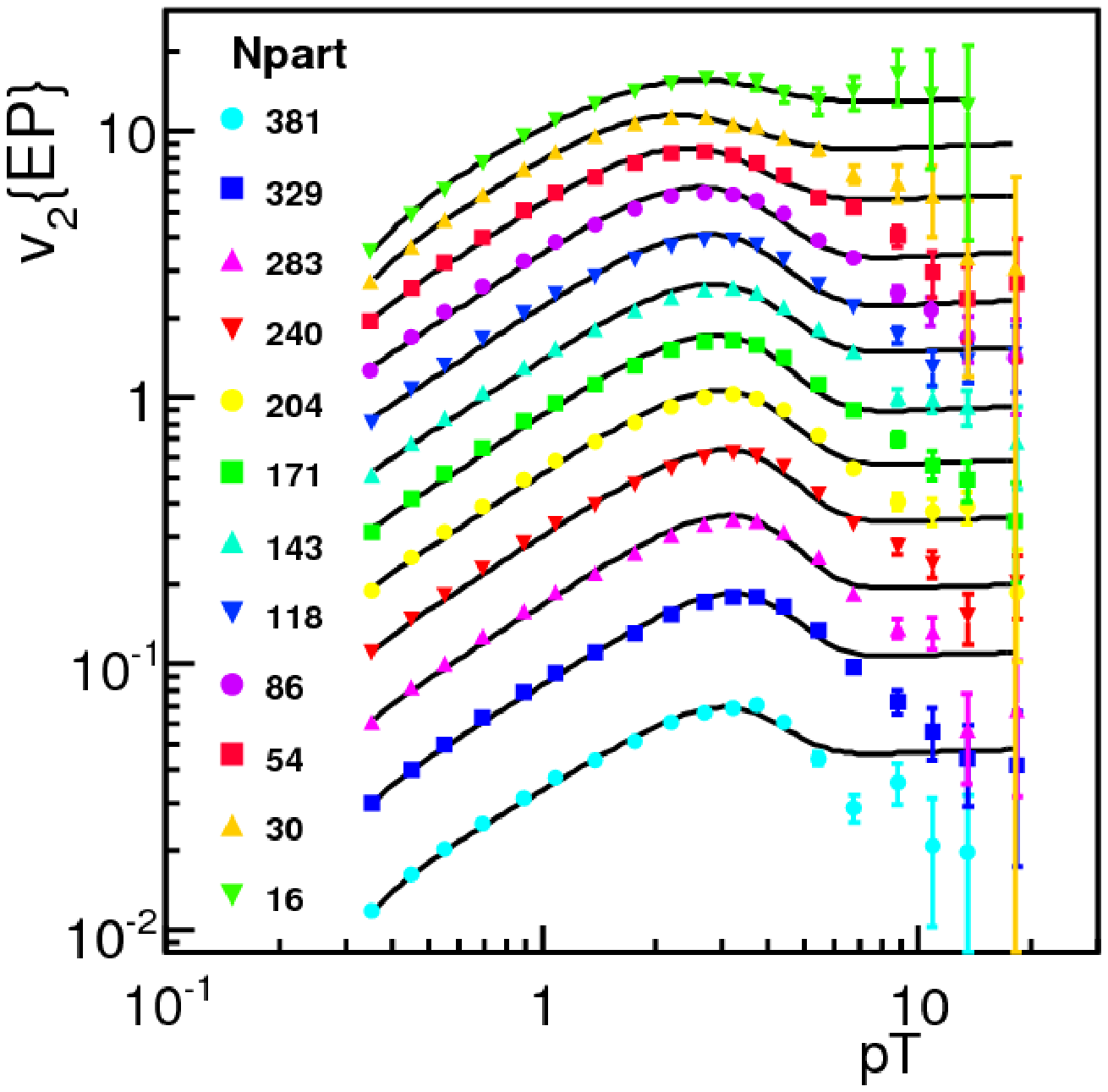} 
\includegraphics[width=0.45\textwidth]{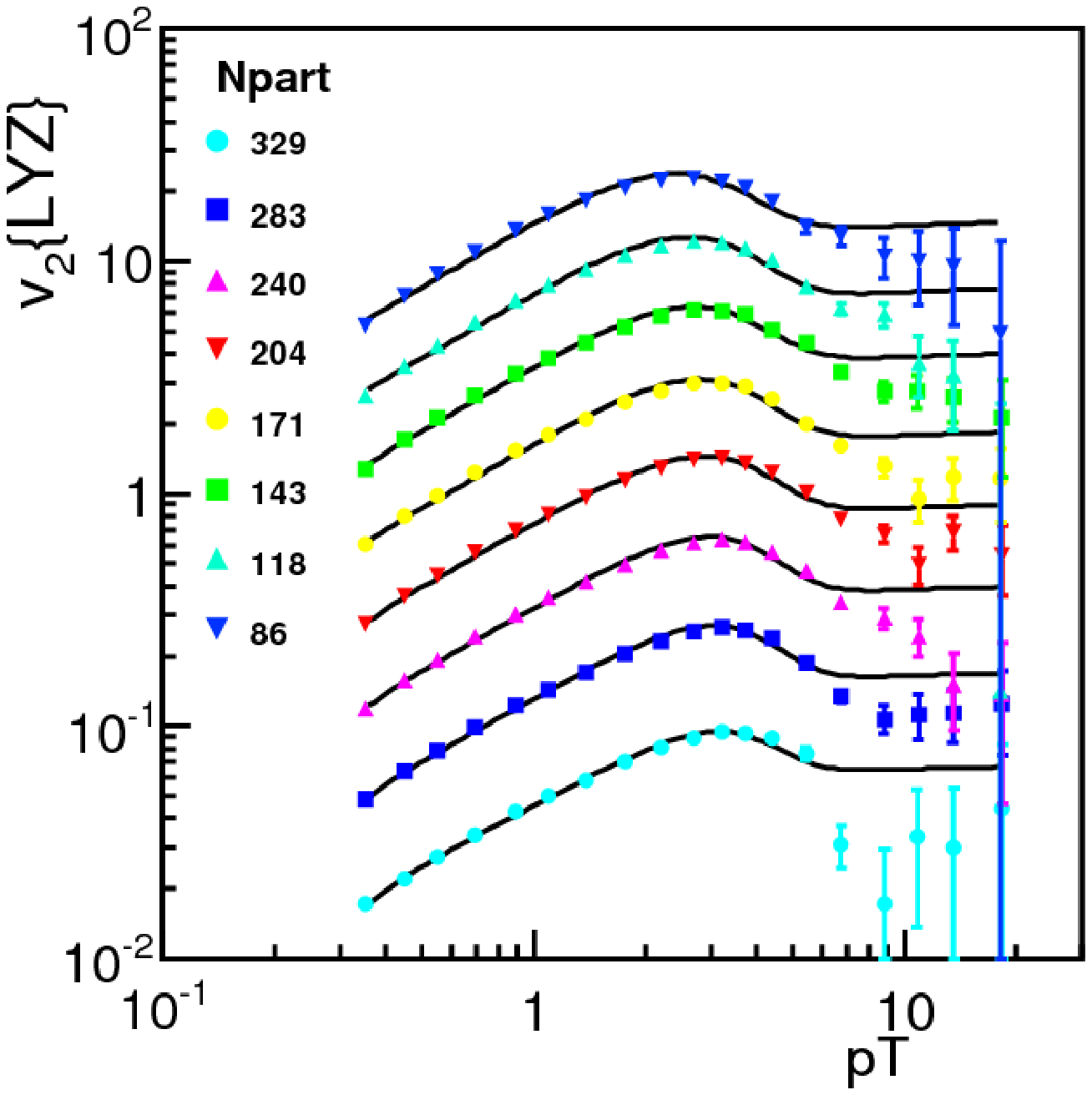} 
\end{center}
\caption{Fits of Eqs.~(\ref{eq17})~--~(\ref{eq18}) to CMS data on $v_2$ \cite{bib:cmsv2} in the case of four types of methods (event plane $v_2\lbrace EP\rbrace$, $2^{nd}$ and $4^{th}$ order cumulant $v_2\lbrace2\rbrace$ and $v_2\lbrace4\rbrace$ and Lee-Yang zeros $v_2\lbrace LYZ\rbrace$ methods). Fit parameters are plotted and enlisted in \cite{bib:UKshAA}.\label{fig:v2}}
\end{figure}

As seen from Sec.~\ref{sec:vn}, up to $\mathcal{O}\left(\delta v^2 \right)$, the transverse spectrum in Eq.~(\ref{eq12}) results in an azimuthal anisotropy of
\be
v_2 = \frac{w_{hard} \,f_{hard} + w_{soft}\, f_{soft}}{f_{hard} + f_{soft}}\;, 
\ee{eq17}
where the coefficient functions are
\be
w_i = \frac{\delta v_i\, \gamma^3_i}{2T_i} \frac{p_T - v_i\, m_T}{1 + \dfrac{q_i-1}{T_i}\,\big[\gamma_i(m_T - v_i\,p_T) - m \big]} \;.
\ee{eq18}
Again, $i=$ soft or hard, $v_i$ are the isotropic part of the transverse flow (denoted by $v_0$ in Eq.~(\ref{eq6}) in Sec.~\ref{sec:vn}). And $\delta v_i$ are the coefficients of $\cos(2\alpha)$ (denoted by $\delta v_2$ in Eq.~(\ref{eq6}) in Sec.~\ref{sec:vn}).

Fits of Eqs.~(\ref{eq17})~--~(\ref{eq18}) to CMS data \cite{bib:cmsv2} on $v_2$ are found in Fig.~\ref{fig:v2}. The four different methods used in \cite{bib:cmsv2} for the extraction of $v_2$ are the $2^{nd}$ and $4^{th}$ order cumulant methods denoted by $v_2\lbrace2\rbrace$ and $v_2\lbrace4\rbrace$, the event-plane $v_2\lbrace EP\rbrace$ and Lee--Yang zeros $v_2\lbrace LYZ\rbrace$ methods. Fitted parameters are listed in \cite{bib:UKshAA}.

Finally, all four methods for the extraction of $v_2$ in \cite{bib:cmsv2} suggest that $\delta v_{soft}$ (the $2^{nd}$ Fourier components of the transverse flow of the soft yields) decreases for more central collisions (see bottom-right panel of Fig.~\ref{fig:dNdpT}). This observation is in accordance with smaller anisotropy in more central collisions.

\section{Summary}
\label{sec:conc}
In this paper, we have \textit{simultanously} reproduced the transverse spectra \textit{and} the azimuthal anisotropy ($v_2$) of charged hadrons stemming from various centrality PbPb collisions at $\sqrt s$ = 2.76 ATeV. In the proposed model, the hadron spectrum is assumed to be simply the sum of yields originated from `soft' and `hard' processes, Eq.~(\ref{eq1}). It is conjectured that hadrons are distributed according to the Tsallis distribution in both types of yields. As for the hard yields, this assumption is supported by the observation that the Tsallis distribution provides a reasonably good approximation for pion spectra obtained via pQCD-improved parton model calculations for central or peripheral PbPb collisions at LHC energy \cite{bib:UKshAA}. Furthermore, the Tsallis distribution describes hadron spectra in pp collisions as well. The soft yields (which we identified by what remains of the hadron spectra after the subtraction of the hard yields) can also be described by a Tsallis distribution with different parameters.

Analytic formulas have been obtained for the spectra and for $v_2$ in the limit of small transverse flow velocity fluctuations as a function of the azimuth angle. The parameters of the soft and hard Tsallis distributions have been determined from fits to transverse spectra and $v_2$ data measured by the CMS \cite{bib:cmsdNdpT,bib:cmsv2} and ALICE \cite{bib:ALICEdNdpT} collaborations. The dependence of the fitted parameters on the event centrality  ($N_{part}$) have been found similar in the case of the CMS and ALICE data. 
Fits to CMS data on $v_2$ suggest that in this model, the anisotropy decreases for more central collisions. Fit parameters are enlisted in \cite{bib:UKshAA}.


\section*{Acknowledgement}
\label{sec:ack}
This work was supported by Hungarian OTKA grants K104260, NK106119, and
NIH TET 12 CN-1-2012-0016. Author GGB also thanks the J\'anos Bolyai
Research Scholarship of the Hungarian Academy of Sciences.


\end{document}